\documentclass[12pt]{article}

\renewcommand{\d}{{\rm d}}
\newcommand{\pOmega}{\,^+\!\Omega}
\newcommand{\mOmega}{\,^-\!\Omega}
\newcommand{\pmOmega}{\,^\pm\!\Omega}
\newcommand{\pR}{\,^{+}\!R}
\newcommand{\mR}{\,^{-}\!R}
\newcommand{\pmR}{\,^{\pm}\!R}
\newcommand{\pv}{\,^{+}\!v}
\newcommand{\mv}{\,^{-}\!v}
\newcommand{\pbv}{\,^{+}\!\mbox{\boldmath$v$}}
\newcommand{\mbv}{\,^{-}\!\mbox{\boldmath$v$}}
\newcommand{\bv}{\mbox{\boldmath$v$}}
\newcommand{\pmbv}{\,^{\pm }\!\mbox{\boldmath$v$}}
\newcommand{\pmv}{\,^{\pm}\!v}
\newcommand{\pmSig}{\,^{\pm}\!\Sigma}
\newcommand{\bl}{\mbox{\boldmath$l$}}
\newcommand{\br}{\mbox{\boldmath$r$}}
\newcommand{\D}{{\cal D}}
\newcommand{\rv}{\mbox{\rm v}}
\newcommand{\sh}{\mathop{\rm sh}\nolimits}

\newcommand{\bomega}{\mbox{\boldmath$\omega$}}
\renewcommand{\Re}{\mathop{\rm Re}\nolimits}

\newcommand{\bSig}{\mbox{\boldmath$\Sigma$}}

\newcommand{\pmbSig}{\,^{\pm}\!\mbox{\boldmath$\Sigma$}}

\newcommand{\bu}{\mbox{\boldmath$u$}}
\newcommand{\be}{\mbox{\boldmath$e$}}
\newcommand{\btau}{\mbox{\boldmath$\tau$}}
\newcommand{\rl}{\ell} 
\newcommand{\pr}{{\hspace{-2.5mm}}^{ \stackrel{\textstyle \prime}{}}}
\newcommand{\prback}{{\hspace{-4mm}}^{ \stackrel{\textstyle \prime}{}}}
\newcommand{\prforw}{{\hspace{-3mm}}^{ \stackrel{\textstyle \prime}{}}}
\newcommand{\N}{{\cal N}}
\renewcommand{\Im}{\mathop{\rm Im}\nolimits}
\newcommand{\bp}{\mbox{\boldmath$p$}}
\newcommand{\bn}{\mbox{\boldmath$n$}}
\newcommand{\overOm}{\overline{\Omega}}
\newcommand{\overT}{\overline{T}}
\begin{document}

\title{On the non-perturbative graviton propagator
}

\author{V.M. Khatsymovsky \\
 {\em Budker Institute of Nuclear Physics} \\ {\em of Siberian Branch Russian Academy of Sciences} \\ {\em
 Novosibirsk,
 630090,
 Russia}
\\ {\em E-mail address: khatsym@gmail.com}}
\date{}
\maketitle

\begin{abstract}
To reduce general relativity to the canonical Hamiltonian formalism and construct the path (functional) integral in a simpler and, especially in the discrete case, less singular way, one extends the configuration superspace, as in the connection representation. Then we perform functional integration over connection. The module of the result of this integration arises in the leading order of the expansion over a scale of the discrete lapse-shift functions and has maxima at finite (Planck scale) areas/lengths and rapidly decreases at large areas/lengths, as we have mainly considered previously; the phase arises in the leading order (Regge action) of the stationary phase expansion.

Now we consider the possibility of confining ourselves to these leading terms in a certain region of the parameters of the theory; consider background edge lengths as an optimal starting point for the perturbative expansion of the theory; estimate the background length scale and consider the form of the graviton propagator. In parallel with the general simplicial structure, we consider the simplest periodic simplicial structure with a part of the variables frozen ("hypercubic"), for which also the propagator in the leading approximation over metric variations can be written in a closed form.
\end{abstract}

keywords: general relativity; piecewise flat spacetime; Regge calculus; discrete connection; functional integral


MSC classes: 83C27; 81S40

\section{Introduction}

Description of general relativity (GR) on a certain class of Riemannian manifolds, namely piecewise flat manifolds, proposed by Regge \cite{Regge}, can be quite self-sufficient, because one can approximate any smooth Riemannian manifold by piecewise flat ones with an arbitrary accuracy \cite{Fein,CMS}. This attracts attention, in particular, in view of the formal non-renormalizability of the continuum GR, on the one hand, and the countability of the number of degrees of freedom of a piecewise flat manifold and the related possibility to apply for quantizing the system the tools used to quantize on the lattice \cite{Ham}, on the other hand. The most convenient and universal for the analysis of quantum Regge gravity is the functional integral approach, in which a certain freedom is in defining the functional measure. Using reasonable physical arguments, it turns possible to fix the measure and use it to obtain physical quantities, such as the Newtonian potential \cite{HamWil1,HamWil2}. In Ref.~\cite{HamLiu}, a diagram technique in simplicial gravity is considered. In Ref.~\cite{RegWil}, a review of the Regge theory and some other approaches is given. Recently the Causal Dynamical Triangulations approach related to the Regge theory has allowed to get interesting results in quantum gravity \cite{cdt}.

In this paper, we consider the question of a perturbative expansion in Regge gravity along with the fixation of the background configuration for it, using the functional integral approach. To construct the functional measure, we use results of our previous papers. One of the approaches to fixing the functional measure might be the canonical Hamiltonian approach, which gives it in the form symbolically as $\d p \d q$ for some set of canonical coordinates $p, q$. It can be given in the original coordinates times some Jacobian of the Poisson brackets of the constraints. In GR, the most convenient is using the tetrad-connection (Cartan-Weyl) form of the action for constructing the Hamiltonian formalism. Especially this refers to the discrete GR or Regge gravity, in which the Regge action in terms of the original edge length variables is quite singular from the viewpoint of passing to the canonical variables.

In the connection representation, the Hamiltonian formalism with area tensors and connection matrices being conjugate variables leads to the Jacobian of the Poisson brackets of the constraints which is singular at the flat metric. This is typical for discrete gravity due to the lack of diffeomorphism invariance, as is reviewed, eg, in Ref.~\cite{Loll}. More convenient is using an analogue of the considered in the literature "area Regge calculus" \cite{areaRC} where the areas of the triangles are independent variables. The equations of motion for this system mean vanishing the defect angles, which, however, because of the lack of a conventional geometric interpretation, does not mean flat spacetime.

In our case, a natural modification consists in considering area {\it tensors} as independent variables \cite{Kha2}. In the configuration superspace of independent area tensors, the points of the physical hypersurface run through sets of area tensors corresponding to all possible sets of the edge vectors. The simple equations of motion ensure the commuting constraints and a simple form of the functional measure, and we must project the measure onto the physical hypersurface. This is achieved by introducing an appropriate delta-function factor with support on the physical hypersurface.

Having constructed the measure, we can make the functional integration over the connection variables and arrive at a functional integral expression in terms of the purely area or lengths variables. In Ref.~\cite{our4}, we have performed the functional integration over connection for the exact selfdual plus anti-selfdual connection representation of the Regge action of the type which we have suggested in Ref.~\cite{our1}. This calculation is done in disregard of the discrete analogs of the Arnowitt-Deser-Misner \cite{ADM} lapse-shift functions. This can serve as a starting point for obtaining the result in the form of a series over (a scale of) the discrete lapse-shift functions, and we have found in Ref.~\cite{our3} master integrals arising in such a calculation.

The result of the functional integration over connection is characterized by the module and by the phase. The leading term of the expansion over the discrete lapse-shift functions gives a pure module and zero phase. On the other hand, we can use the stationary phase method and expand the action as a function of the connection around the classical solution. This classical solution for the connection just leads to the Regge action by construction of the connection representation. Thus, the leading order of the stationary phase expansion of the result of the integration gives a pure phase, Regge action, and the trivial module.

Usually the most confident result for any calculated value in a diagram technique for a path integral is the one that appears already in the leading order. Therefore, it is expedient to represent the module of the result of the functional integration over connection in the form of the expansion over the lapse-shift functions and its phase in the form of the stationary phase expansion. These expansions go over different parameters, but upon some analysis it turns out that there is a region of the parameters of the theory where both the expansion parameters are small and we can confine ourselves to the leading terms in them.

With the functional integral in terms of lengths variables, we arrive at the task to construct a perturbation series if the functional measure is bell-shaped. Defining the optimal background lengths in such an expansion is close to finding the measure maxima; also the background length scale is estimated and typical features of the propagator are noted.

We aim at analyzing the system on the general simplicial structure, but for purposes of illustrating and modeling, we simultaneously consider the action also on the simplest periodic Regge manifold (with a hypercubic cell) and reduce the set of variables by freezing some variables or their combinations. Upon excluding the connection, the resulting "hypercubic" action tends to the GR action for the slowly varying metric from site to site (the continuum limit). In this model, we have the same mechanism of the background length scale fixing; we can also write out the propagator in a closed form.

In the next Section, the mentioned selfdual plus anti-selfdual form of the action and the associated notation are given, the hypercubic action introduced, and the conditions on the parameters of the theory are considered that the result of the functional integration over connection can be approximated by the already known leading terms in the module (earlier found) and in the phase (the action with the classically excluded connection). In Section \ref{tetrad}, the remaining, tetrad part of the measure is taken into account and the full resulting measure after the integration is considered, including the hypercubic case. In Section \ref{perturbation}, a condition on the starting point (background lengths) for the perturbative expansion is found and the background length scale is estimated, including the hypercubic case. In Section \ref{propagator}, we discuss typical features of the propagator at the background length scale and give its form for the hypercubic system for the case when the tetrad is slowly varying from site to site. Then Conclusions are made.

\section{Action and functional integration over connection}

Here we write out the selfdual plus anti-selfdual form of the action (\ref{S-simplicial}) (which we used earlier) and give the associated notation, consider the action for the simplest periodic Regge manifold on a reduced set of variables (the "hypercubic" action) (\ref{S-cube}) and its relation to the GR action in the continuum limit, write out the earlier obtained module of the result of the functional integration over connection (\ref{N0}) and specify it also in the hypercubic case, (\ref{N0-cube}), briefly consider the stationary phase expansion for the phase and conditions on the parameters of the theory (of the type (\ref{a-1-e-a})) that the result of the functional integration could be approximated by the already known leading terms in the module (earlier found) and in the phase (the action with the classically excluded connection).

\subsection{General simplicial action and notation}

The piecewise flat spacetime is considered as the simplicial complex consisting of the 4-dimensional tetrahedra or 4-simplices $\sigma^4$, their 3-dimensional faces $\sigma^3$ (usual tetrahedrons), 2-dimensional faces or triangles $\sigma^2$, edges $\sigma^1$ and vertices $\sigma^0$. Different variables can be given on these objects. We assign a local pseudo-Euclidean frame to each 4-simplex. There are SO(3,1) connection matrices $\Omega_{\sigma^3}$ on the 3-faces $\sigma^3$ and curvature matrices $R_{\sigma^2}$ on the triangles $\sigma^2$. The discrete analogs of connection and curvature were introduced by Fr\"{o}hlich \cite{Fro}. The matrix $R_{\sigma^2}$ is the holonomy of $\Omega_{\sigma^3}$, a path ordered product of $\Omega_{\sigma^3}, \sigma^3 \supset \sigma^2$, taken along the loop enclosing $\sigma^2$. There are vectors $l^a_{\sigma^1}$ of the edges $\sigma^1$ and area tensors $v^{ab}_{\sigma^2}$ of the triangles $\sigma^2$. The matrices $\Omega_{\sigma^3}$ (and $R_{\sigma^2}$) can be decomposed multiplicatively into the selfdual $\pOmega_{\sigma^3}$ and anti-selfdual $\mOmega_{\sigma^3}$ parts (accordingly $\pR_{\sigma^2}$ and $\mR_{\sigma^2}$), elements of SO(3,C). The matrices $v^{ab}_{\sigma^2}$ decompose into selfdual $\pv^{ab}_{\sigma^2}$ and anti-selfdual $\mv^{ab}_{\sigma^2}$ parts additively.

The action $\frac{1}{2} \int R \sqrt{-g} \d^4 x$ on the piecewise flat manifold in the representation which we use in the paper \cite{our4} takes the form
\begin{eqnarray}\label{S-simplicial}                                        
S ( v, \Omega ) = \frac{1}{2} \sum_{\sigma^2} \left [ \left ( 1 + \frac{i}{\gamma } \right ) \sqrt{ \pbv^2_{\sigma^2} } \arcsin \frac{\pbv_{\sigma^2} * \pR_{\sigma^2} ( \Omega )}{\sqrt{ \pbv^2_{\sigma^2}}} \right. \nonumber \\
\left. + \left ( 1 - \frac{i}{\gamma } \right ) \sqrt{ \mbv^2_{\sigma^2} } \arcsin \frac{\mbv_{\sigma^2} * \mR_{\sigma^2} ( \Omega )}{\sqrt{ \mbv^2_{\sigma^2}}} \right ].
\end{eqnarray}

\noindent Here, $\bv * R \equiv \frac{1}{2}v^i R^{kl} \epsilon_{ikl}$, $\epsilon_{123} = +1$, and $\pmbv$ (or $\pmv_k$ in components) are 3-vectors parameterizing the selfdual $\pv^{ab}$ and antiselfdual $\mv^{ab}$ parts of the antisymmetric tensor $v^{ab}$ expanded over triple of (anti-)selfdual basis matrices $\pmSig_k$ obeying algebra of the Pauli matrices times $-i$,
\begin{eqnarray}                                                            
& & v^{ab} \! = \! \pv^{ab} \! + \! \mv^{ab}, \pmv^{ab} \! = \! \frac{1}{2} v^{ab} \! \pm \! \frac{i}{4}
\epsilon^{ab}_{~~cd}v^{cd} \! = \! \pmv^k \pmSig^{ab}_k/ 2, (\pv^{ab})^* \! = \! \mv^{ab}, ~ \epsilon_{abcd} \! \pmbSig^{cd} \nonumber \\
& & = \! \mp 2 i \pmbSig_{ab}, \pmSig_{kab} \! = \! -\epsilon_{kab} \! \pm \! i(g_{ak}g_{0b} \! - \! g_{a0}g_{kb}), \pmSig_k \! \pmSig_l \! = \! - \! \delta_{k l} \! + \! \epsilon_{kl}{}^m \! \pmSig_m.
\end{eqnarray}

\noindent Here $\epsilon^{0123} = +1$, the metric $g_{ab} ={\rm diag}(-1,1,1,1)$. In particular, if $v^{ab}$ is the area bivector built on some edge vectors $l^a_1$ and $l^a_2$, $v^{ab} = \frac{1}{2}\epsilon^{ab}{}_{cd}l^c_1l^d_2$, then
\begin{equation}                                                            
2 \pmbv = -i l^a_1 l^b_2 \pmbSig_{a b} = \pm i \bl_1 \times \bl_2 - \bl_1 l^0_2 + \bl_2 l^0_1.
\end{equation}

\noindent In what follows, we usually consider the selfdual SO(3,C) vectors and omit "+" on them; the anti-selfdual ones follow by complex conjugation. In $S ( v, \Omega )$, $\pmOmega$ and $\pmR$ act as SO(3,C) matrices $\pmOmega^k{}_l$ and $\pmR^k{}_l$. A rotation $\pOmega^k{}_l$, which has a generator $\epsilon^{km}{}_l \omega_m $ (adjoint representation), is equivalent to two selfdual rotations over the local spacetime indices with the generator $\bomega \bSig / 2$ (fundamental representation),
\begin{equation}                                                            
\exp (\bomega \bSig / 2) \Sigma_l \exp ( - \bomega \bSig / 2) = \Sigma_k \pOmega^k{}_l , ~~~ \pOmega^k{}_l = ( \exp \| \epsilon^{im}{}_l \omega_j \| )^k{}_l.
\end{equation}

\noindent The quantity $\gamma$ is a discrete analog of the Barbero-Immirzi parameter \cite{Barb,Imm}, which parameterizes the parity odd Holst term \cite{Holst,Fat} in the continuum theory.

We adopt a partly regular structure of the 4-dimensional simplicial complex. There are 3-dimensional leaves of the foliation, simplicial complexes of the same structure numbered by an integer coordinate $t$. The 4-dimensional geometry is constructed by connecting analogous vertices in the neighboring leaves by edges. We call these edges {\it $t$-like}. Besides that, these leaves are connected by {\it diagonal} edges. A diagonal edge connects a vertex $\sigma^0_1$ in one leaf and a vertex $\sigma^0_2$ in a neighboring leaf analogous to a vertex in the former leaf neighboring to $\sigma^0_1$. A {\it leaf} edge is completely contained in a leaf. A $t$-like simplex contains a $t$-like edge, a leaf simplex is completely contained in a leaf, and any other simplex will be called a diagonal one.

The region between any two neighboring leaves gets divided into 4-dimensional prisms whose lateral 3-dimensional surface is formed by $t$-like tetrahedra $\sigma^3$; the bases of each prism are analogous 3-simplices in these leaves, and each prism is divided into four 4-simplices. The vectors of the $t$-like edges are discrete analogs of the Arnowitt-Deser-Misner \cite{ADM} lapse-shift functions if $t$ is considered as a time coordinate. The lapse-shift functions in the continuum theory can be considered as gauge parameters, the fixation of which means fixing four degrees of freedom in the metric tensor associated with diffeomorphisms.

In this construction, any given 4-simplex $\sigma^4$ contains a $t$-like edge with a 4-vector $l^a_0$, discrete lapse-shift functions, and three else edges with 4-vectors $l^a_\alpha$, $\alpha = 1, 2, 3$, with a common vertex $O$. The tetrad $l^a_\lambda$, $\lambda = 0, 1, 2, 3$, forms six bivectors
\begin{equation}\label{v=ll}                                                
v^{ab}_{\lambda\mu} \equiv \frac{1}{2} \epsilon^{ab}{}_{cd} l^c_{\lambda} l^d_{\mu}, ~~~ v^{ab}_1 \equiv  v^{ab}_{2 3}, ~ 2 ~ \mbox{perm}(123) , ~~~ \tau^{ab}_\alpha \equiv v^{ab}_{0 \alpha}.
\end{equation}

\noindent Three of them $v^{ab}_\alpha$ are bivectors of some leaf/diagonal triangles, and three $\tau^{ab}_\alpha$ are bivectors of some $t$-like ones.

\subsection{The action for the simplest periodic Regge manifold and a reduced set of variables}\label{hypercubic}

In parallel with the action (\ref{S-simplicial}) on the above rather general simplicial structure, we consider this action on the simplest periodic simplicial system with a part of the degrees of freedom frozen and with the vertices (sites) at the integer coordinates $x^\lambda$,
\begin{eqnarray}\label{S-cube}                                              
S_{\rm cube} ( v, \Omega ) = \frac{1}{4} \sum_{\stackrel{\scriptstyle \lambda \mu \nu \rho}{\rm sites}} \left ( 1 + \frac{i}{\gamma } \right ) \epsilon^{\lambda \mu \nu \rho} \sqrt{ \bv^2_{\lambda \mu} } \arcsin \frac{\bv_{\lambda \mu} * \pR_{\nu \rho} ( \Omega )}{\sqrt{ \bv^2_{\lambda \mu}}} + {\rm c. c. },
\end{eqnarray}

\noindent where "c.c." means "complex conjugate" and
\begin{equation}\label{v=ll-R=wwww}                                         
2 \bv_{\lambda\mu} = i \bl_\lambda \times \bl_\mu + l^0_\lambda \bl_\mu - l^0_\mu \bl_\lambda, ~~~ R_{\lambda\mu} (\Omega ) = \overOm_\lambda (\overT_\lambda \overOm_\mu) (\overT_\mu \Omega_\lambda) \Omega_\mu,
\end{equation}

\noindent $T_\lambda$ ($\overline{T}_\lambda$) is the shift operator from $x^\lambda$ to $x^\lambda + 1$ ($x^\lambda - 1$).

This action can be obtained from $S ( v, \Omega )$ on the simplest periodic Regge manifold with the hypercubic cell divided by diagonals into 24 4-simplices, by freezing some of (combinations of) the variables $ v, \Omega $ (this manifold was considered, eg, in Ref. \cite{RocWilPL}). Namely,

1) $\Omega_{\sigma^3} = 1$ for the internal $\sigma^3$s in the 4-cube,

2) $\Omega_{\sigma^3} = \Omega_\lambda$ for each of the six $\sigma^3$ into which the 3D face (the 3-cube orthogonal to the edge along $x^\lambda$) is divided,

3) the area tensors of two triangles constituting a quadrangle in the $x^\lambda, x^\mu$ plane are the same.

\noindent This can be referred to as "hypercubic action", although this is a "mini-superspace-in-mini-superspace" simplicial action.

Excluding $\Omega$ from the action $S ( v, \Omega )$ with the help of the equations of motion, we get the Regge action. For $S_{\rm cube} ( v, \Omega )$, an action $S_{\rm cube} ( v, \Omega_0 (v ) )$ tends to the GR action for a slowly varying metric from site to site ($\Omega_0 (v )$ is a particular solution of the equations of motion for $\Omega$). Indeed, assuming that $\bomega_\lambda$ for the solution of the equations of motion for $\Omega_\lambda$ ($\Omega_\lambda \equiv \exp \omega_\lambda$) are small (which is confirmed by the subsequent calculation), we can expand over them,
\begin{eqnarray}\label{r}                                                   
& & 2 r^k_{\lambda \mu} \equiv - \epsilon^k{}_{l m} \pR^{l m}_{\lambda \mu}, ~~~ \br^{(0)}_{\lambda \mu} = \delta_\lambda \bomega_\mu - \delta_\mu \bomega_\lambda + \bomega_\lambda \times \bomega_\mu, ~~~ \br_{\lambda \mu} = \br^{(0)}_{\lambda \mu} - \bomega_\lambda \times \delta_\lambda \bomega_\mu \nonumber \\ & & - \delta_\mu \bomega_\lambda \times \bomega_\mu+ \frac{1}{2} \bomega_\lambda \times \delta_\mu \bomega_\lambda + \frac{1}{2} \delta_\lambda \bomega_\mu \times \bomega_\mu - \frac{1}{2} \delta_\lambda \bomega_\mu \times \delta_\mu \bomega_\lambda + O ( \omega^3 ).
\end{eqnarray}

\noindent Here $\delta_\lambda = 1 - \overT_\lambda$. Keeping in (\ref{S-cube}) the linear and quadratic in $\omega$ terms and taking $\br_{\lambda \mu} = \br^{(0)}_{\lambda \mu}$ we get in the leading approximation over $\delta$ and $\omega$
\begin{equation}\label{r(0)}                                                
- \frac{1}{4} \sum_{\stackrel{\scriptstyle \lambda \mu \nu \rho}{\rm sites}} \left ( 1 + \frac{i}{\gamma } \right ) \epsilon^{\lambda \mu \nu \rho} \bv_{\lambda \mu} \cdot \br^{(0)}_{\nu \rho} + {\rm c. c. },
\end{equation}

\noindent which is a finite difference form of the continuum connection representation \cite{Holst,Fat} of the action $\frac{1}{2} \int R \sqrt{-g} \d^4 x$,
\begin{eqnarray}\label{Cartan}                                             
& & S ( e, \omega ) = - \frac{1}{8}\int{(\epsilon_{abcd}e^a_{\lambda}e^b_{\mu} + \frac{2}{ \gamma}e_{\lambda c}e_{\mu d })\epsilon^{\lambda\mu\nu\rho} [\partial_{\nu} + \omega_{\nu}, \partial_{\rho} + \omega_{\rho}]^{cd}{\rm d}^4x} = - \frac{1}{8} \epsilon^{\lambda\mu\nu\rho} \cdot \nonumber \\ & & \hspace{-2mm} \cdot \left ( 1 + \frac{i}{\gamma } \right ) \int (i \be_\lambda \times \be_\mu + e^0_\lambda \be_\mu - e^0_\mu \be_\lambda) \cdot (\partial_\nu \bomega_\rho - \partial_\rho \bomega_\nu + \bomega_\nu \times \bomega_\rho) \d^4 x + {\rm c. c. },
\end{eqnarray}

\noindent where according to the relationship between the continual tetrad $e^a_\lambda$ and the shift vector by $\Delta x^\lambda$, $l^a = e^a_\lambda \Delta x^\lambda$, and $\Delta x^\lambda = 1$ for the neighboring vertices, $l^a_\lambda$ can be identified with $e^a_\lambda$. Excluding $\omega$ from the leading order form (\ref{r(0)}) gives $\omega = O( \delta )$, and taking into account the corrections to $\br^{(0)}_{\lambda \mu}$ in $\br_{\lambda \mu}$ (\ref{r}) and the corresponding corrections to (\ref{r(0)}) of higher orders in $\delta$.

There is a more direct correspondence between the connection action on a periodic simplicial complex and the form on the hypercubes (\ref{r(0)}) when approaching the continuum limit. This can be seen on the simplest example of the hypercubic cell divided by diagonals into 24 4-simplices. The area tensors of the triangles contained in the cell can be expressed with the required accuracy as combinations of the "elementary" ones $v^{a b}_{\lambda \mu}$ in the flat approximation. Quite expectedly, and we have checked in \cite{Kha4} that the quadratic form of the connection in $S ( v, \Omega )$ reduces with the same accuracy to (\ref{r(0)}) with $\omega_\lambda$ being a combination of $\omega_{\sigma^3}$ in the cell,
\begin{equation}\label{omega-lambda=sum-omega-sigma}                       
\omega_\lambda = \sum_{\sigma^3 \subset \mbox{cell}} \prforw \pm \omega_{\sigma^3},
\end{equation}

\noindent an algebraic sum over $\sigma^3$ encountered when we pass through the cell in the $x^\lambda$ direction.

\subsection{The result of the functional integration over connection}

The functional integration of $\exp [i S ( e, \omega )]$ over the connection $\omega$ in the continuum theory is Gaussian and gives $\exp{ (i \frac{1}{2} \int R \sqrt{-g} \d^4 x ) }$ in the tetrad $e$ variables.

In the discrete theory, we use the expansion in the lapse-shift functions for the module of the result and the stationary phase expansion for its phase, as noted in Introduction.

The integration measure $\D \Omega$ (the product of the invariant or Haar measures on the instances of SO(3,1) group) contains the product of independent integrations over $\D R_{\sigma^2}$ for the leaf and diagonal triangles $\sigma^2$. The holonomy on the $t$-like triangles can be written as (multiplicative) expressions in terms of the holonomy on the leaf/diagonal triangles and connection on the leaf/diagonal tetrahedra (these expressions just solve some algebraical identities, the Bianchi identities, \cite{Regge}). This underlies the expansion over the lapse-shift functions. The action reads
\begin{eqnarray}\label{leaf/diag+t-like}                                   
& & S ( v, \Omega ) = \frac{1}{2} \left ( 1 + \frac{i}{\gamma } \right ) \left [ \sum_{\stackrel{{\scriptstyle\rm leaf/dia-}}{{\rm gonal~}\sigma^2}} \sqrt{ \bv^2_{\sigma^2} } \arcsin \frac{\bv_{\sigma^2} * \pR_{\sigma^2}}{\sqrt{ \bv^2_{\sigma^2}}} \right. \nonumber \\ & & \hspace{-10mm} \left. + \sum_{t {\rm -like~}\sigma^2} \sqrt{ \btau^2_{\sigma^2} } \arcsin \frac{\btau_{\sigma^2} * \pR_{\sigma^2} ( \{ \pR_{\sigma^2} | {\scriptstyle \rm leaf/diagonal~} \sigma^2 \} )}{\sqrt{ \btau^2_{\sigma^2}}} \right ] + {\rm c. c. }
\end{eqnarray}

\noindent The module of the result of integration over $\D \Omega$ in the leading order $\btau = 0$, as we have found in \cite{our4}, is the product of some functions $\N_0 (\rv_{\sigma^2} ) = \N_0 (\rv_{\sigma^2}, \rv^*_{\sigma^2} )$ over the leaf/diagonal triangles $\sigma^2$, $\rv = \sqrt{\bv^2}$,
\begin{equation}\label{N0}                                                 
\N_0 (\rv ) = \left | \frac{1}{\frac{1}{4} \left ( \frac{1}{\gamma} - i \right )^2 \rv^2 + 1} \frac{\frac{1}{4} \left ( \frac{1}{\gamma} - i \right ) \rv}{ \sh \left [ \frac{\pi}{2} \left ( \frac{1}{\gamma} - i \right ) \rv \right ]} \right |^2 .
\end{equation}

\noindent Master integrals (with the $n$-th order monomials of curvature variables inserted) \cite{our3} behave as $| \rv |^{- n}$ relative to $\N_0$ at large $| \rv |$ and define the $n$-th order correction $\sim (|\btau | |\rv |^{- 1})^n \sim (\varepsilon l^{- 1})^n$ where $| \btau |$ is a $t$-like area vector scale, $\varepsilon$ is a scale of the lapse-shift vectors, $l$ is a scale of the leaf/diagonal edge lengths.

The expansion for the module (even orders) is in powers of $(| \btau | | \rv |^{- 1})^2 \sim \varepsilon^2 l^{- 2} $.

The $S_{\rm cube}$ looks as a sum over a set of pairwise equal triangles with the area vectors $\bv_{\lambda \mu}$ and $\bv_{\mu \lambda}$ and equal contributions or as a sum over areas (quadrangles) with the area vectors $2 \bv_{\lambda \mu}$ with, say, $\lambda > \mu$. The module of the result of integration over $\D \Omega$ is the same, but refers to the quadrangles of the 3D space sections with area vectors $2 \bv_\alpha, \alpha = 1, 2, 3$,
\begin{equation}\label{N0-cube}                                            
\prod_{\mbox{sites}} \prod_\alpha \N_0 (2 \rv_\alpha ).
\end{equation}

\noindent

Now consider the stationary phase expansion. We expand the action as a function of (the generator of) the connection $\omega$ in Taylor series around the classical solution and develop the perturbation theory for the functional integral with respect to the quadratic form of the connection $(\omega B \omega)$,
\begin{equation}\label{stationary}                                         
\int \exp [ i ( S(v, \Omega_0 ) + (\omega B \omega) + O( \omega^3 ) ) ] D \omega = \exp [ i ( S(l )] \int \exp [ i (\omega B \omega) ] ( 1 + O( \omega^3 ) ) D \omega ,
\end{equation}

\noindent where the connection $\Omega = \Omega_0 \exp \omega $, $\omega^{ab} = - \omega^{ba}$, and $\Omega_0 (v )$ is a particular solution of the equations of motion for $\Omega$ so that $S(v, \Omega_0 (v ) )$ is the Regge action $S(\rl )$ (a function of the edge lengths $\rl$) and the linear in $\omega$ term is absent.

In the case of $S_{\rm cube} ( v, \Omega )$, as considered in Subsection \ref{hypercubic}, the leading term $S_{\rm cube} (\rl ) \equiv S_{\rm cube} ( v, \Omega_0 (v ) )$ is an action, which tends to the GR action for a slowly varying metric from site to site.

Eq. (\ref{stationary}) is a series of integrals that are similar to Gaussian integrals.

One of the deviations from the Gaussian type is in the integration limits. They are determined by the fact that the element of SO(3,C), parameterized by $\bomega$, is a rotation by the complex angle $\sqrt{\bomega^2} \mbox{mod} (2 \pi )$. That is, the actual integration region should be the band of the complex plane of $\sqrt{\bomega^2}$, $0 \leq \Re \sqrt{\bomega^2} < 2 \pi$. The integral over this region is the integral with the infinite limits plus corrections, which are powers of $\tilde{\omega } / (2 \pi )$ in relative magnitude. Here $\tilde{\omega }$ is a typical connection dominating in the integral. At the same time, the relative values of the subsequent terms in the expansion are powers of $\tilde{\omega }$. Since we aim at the situation when the subsequent terms can be neglected in comparison with the leading term (Regge action), the integration limits can be considered infinite as well.

The remaining deviation from the Gaussian type is in the measure $ D \omega$. It is the product over $\sigma^3$s of the Haar measures on $\mbox{SO(3,C)} \times \mbox{SO(3,C)}$ of the type $\D \bomega \D \bomega^*$ (it is assumed that $\d^3 \bomega \d^3 \bomega^* \equiv 2^3 \d^3 \Re \bomega \d^3 \Im \bomega$ for the Lebesgue integration element). We can pass to a new complex 3-vector variable $\bu = \bomega + O( \bomega^3 )$ so that this measure would be Lebesgue one. Then we rename $\bu$ back to $\bomega$. The measure $\D \bomega$ differs from $\d^3 \bomega$ by a factor, a function of the spherical radius $\sqrt{\bomega^2 }$. It is sufficient to redefine the latter leaving the angle coordinates the same. This looks as a modification of the expression $\Omega = \exp \omega$. As a result, the terms $O( \omega^4 )$ and of higher orders in the exponent are modified numerically, but not parametrically.

Consider possible parameters of the stationary phase expansion of the phase of interest and its compatibility with the expansion of the module of interest over the lapse-shift functions. The form $(\omega B \omega)$ contains the terms $\bomega_{\sigma^3_1} \times \bomega_{\sigma^3_2} \cdot \btau_{\sigma^2}$ provided by the $t$-like triangles $\sigma^2$ and $\bomega_{\sigma^3_1} \times \bomega_{\sigma^3_2} \cdot \bv_{\sigma^2}$ provided by the leaf/diagonal $\sigma^2$. In $\bomega_{\sigma^3_1} \times \bomega_{\sigma^3_2} \cdot \btau_{\sigma^2}$, both $\sigma^3_1$ and $\sigma^3_2$ are $t$-like ($\bomega_\alpha \times \bomega_\beta \cdot \btau_\gamma \epsilon^{\alpha \beta \gamma} , \btau_\alpha \equiv \bv_{0 \alpha}$ in the hypercubic model). The terms $\bomega_{\sigma^3_1} \times \bomega_{\sigma^3_2} \cdot \bv_{\sigma^2}$ for the leaf/diagonal $\sigma^2$s provide all three combinations for the pair of the tetrahedra $\sigma^3_1$ and $\sigma^3_2$: both are $t$-like, as in the $\bomega \bomega \btau$ terms ($\overline{T}_0 \bomega_\alpha \times \bomega_\alpha \cdot \bv_\alpha , \bv_\alpha \equiv \epsilon_{\alpha \beta \gamma} \bv_{\beta \gamma} / 2$ in the hypercubic model), $t$-like and leaf/diagonal, and both are leaf/diagonal. The expression $\det B$ is not reducible to a product of local expressions, and there is no reason for its being identically zero at $\tau = 0$; $\det B (\tau = 0) \neq 0$ for a random configuration. This means that the typical values of $\omega_{\sigma^3}$ in the integral (\ref{stationary}) at small $| \btau |$ are defined by the $\bomega \bomega \bv$ terms and are equal to $\rv^{ - 1 / 2 } = 1 / l$, and, by parity, the stationary phase expansion is over $1 / l^2$.

For certain configurations, $\det B (\tau = 0) = 0$, and the expansion terms over $1 / l^2$ are singular, then the $\bomega \bomega \btau$ terms play a regularizing role and lead to the fact that the potentially infinite typical values of certain components of $\omega_{\sigma^3}$ in the integral are of the order of $O ( | \btau |^{- 1 / 2} )$, and the expansion contains powers of $| \btau |^{- 1 / 2} = ( \varepsilon l)^{ - 1 / 2}$ or, by parity, powers of $1 / ( \varepsilon l )$.

In what follows, in the course of subsequent functional integration over tetrad variables for finding physical quantities, the scale $l$ (a dynamical variable) will be replaced by a constant $a$. Since a singularity in the phase of a complex integrand is integrable, a correction $O(l^{- 2})$ to the action will result in a relative correction $O(a^{- 2})$ to the physical result. This parameter $a^{-2}$ and that of the module expansion at $l = a$, $\varepsilon^2 a^{- 2}$, can be both small and allow us to limit ourselves to the leading terms in the expansions, at
\begin{equation}\label{e-a}                                                
a >> 1, ~ \varepsilon << a.
\end{equation}

\noindent It is safer to foresee applications sensitive to singularities and assume that the expansion is over $1 / ( \varepsilon a )$. The parameters of it and of the module expansion, $\varepsilon^2 a^{- 2}$, can be both small and allow us to limit ourselves to the leading terms in the expansions, at
\begin{equation}\label{a-1-e-a}                                            
a >> 1, ~ a^{-1} << \varepsilon << a.
\end{equation}

\section{Tetrad part of the functional measure}\label{tetrad}

Now consider the rest, except for $\D \Omega$, in the full tetrad-connection functional measure or the full measure itself in more detail. In some previous papers we touched on this point; now we are interested in the order of the measure with respect to the area scale per triangle and single out measures on separate areas ((\ref{Nv11dv}), including the module of the result of the functional integration over $\D \Omega$ (\ref{N0})) or estimate the area scale dependence of the full measure on the tetrad type variables $F$ in the approximation of the factorization over triangles (\ref{F(l)dnl}). The hypercubic version $F_{\rm cube}$ is written, (\ref{Fcube(l)dnl}), its (combinatorial-topological) difference from $F$ is considered.

As mentioned in Introduction, to define the functional measure, we consider area tensors as independent variables, then project the measure onto the physical hypersurface in the configuration superspace. A source of the functional measure from the first principles is the canonical Hamiltonian formalism leading to $\d p \d q$ for conjugate variables $p, q$ with the standard kinetic term $p \dot{q}$. This formalism is defined in the continuous time limit when the distances and $\Delta t$ between the neighboring leaves are arbitrarily small. For the corresponding kinetic term ${\rm tr} ( v \Omega^\dagger \dot{ \Omega } )$ (or, more exactly, its combination with ${\rm tr} ( ^*v \Omega^\dagger \dot{ \Omega } )$, $^*v^{ab} \equiv \epsilon^{ab}_{~~cd}v^{cd} / 2$) we can get the measure symbolically as (a product of)
\begin{equation}\label{dvDOmega}                                           
\d^6 v \D \Omega ,
\end{equation}

\noindent where $\D \Omega$ is the Haar measure. (Here $v \equiv v_{\sigma^2}$ and $\Omega \equiv \Omega_{\sigma^3}$ for certain pairs $\sigma^2$ and infinitesimal $\sigma^3$ having $\sigma^2$ as its base, and $v_{\sigma^2} \equiv v_{\sigma^2 | \sigma^4}$ is defined in the frame of a certain $\sigma^4$ containing $\sigma^2$.) We can go backward to the usual (not shrunk in the $t$ direction) simplicial complex and write down the measure on the area tensors and connection which would result in the canonical formalism measure if any direction is taken as a time and the continuous time limit is taken. Roughly, it is intended to be the product of $\d^6 v_{\sigma^2 | \sigma^4}$ for all $\sigma^2$, $\sigma^4 \supset \sigma^2$ and $\D \Omega_{\sigma^3}$ for all $\sigma^3$. More exactly, first, four values related to the $t$-like area tensors should be fixed in order that this would result in fixing four components of the $t$-like vector at each vertex, the analogs of the lapse-shift functions, when passing to the physical hypersurface of Regge calculus. For example, considering the tetrad and bivectors (\ref{v=ll}) chosen at the given vertex $\sigma^0$ and defined in some 4-simplex as functions of this vertex, we can take the following four conditions,
\begin{equation}\label{vv=e2}                                              
\tau_\alpha ( \sigma^0 ) \circ \tau_\alpha ( \sigma^0 ) = \tilde{\varepsilon}^2, ~~~ \alpha = 1, 2, 3, ~~~ \tau_1 ( \sigma^0 ) \circ \tau_2 ( \sigma^0 ) = 0, ~~~ A \circ B \equiv A_{ab} B^{ab} / 2.
\end{equation}

\noindent A certain feature of these conditions is that if $l^0_\alpha ( \sigma^0 ) = 0$, $\alpha = 1, 2, 3$ (Schwinger time gauge \cite{Schw}), and $\bl_\alpha ( \sigma^0 ) \cdot \bl_\beta ( \sigma^0 ) \propto \delta_{\alpha \beta}$ is given, then (\ref{vv=e2}) fixes lapse $l^0_0 ( \sigma^0 ) \propto \tilde{\varepsilon}$ and shift $\bl_0 ( \sigma^0 ) = 0$. Second, there can be integrations only for six out of ten triangles in the 4-simplex - the four others are algebraic sums of these six. As such, we can take the six triangles that contain the above common vertex $O$ for the tetrad (mentioned when considering (\ref{v=ll})). Once we accept the above described partly regular structure, where each 4-simplex $\sigma^4$ contains a $t$-like edge, we take the common vertex $O$ as the initial (past) end point of the $t$-like edge. There are three $t$-like and three leaf/diagonal triangles for which the integrations are introduced. In the physical case, these six bivectors are parameterized in terms of the discrete tetrad 4-vectors (\ref{v=ll}). Let the prime on the product over triangles in the 4-simplex means the restriction to the independent six triangles. Thus, the measure to be further projected onto the physical hypersurface is
\begin{eqnarray}\label{area-tensor-measure}                                
\prod_{\sigma^0}{\left (\delta (\tau_1 (\sigma^0) \circ
\tau_2 (\sigma^0)) \prod^3_{\alpha=1} {\delta
(\tau_\alpha (\sigma^0) \circ \tau_\alpha (\sigma^0) - \tilde{\varepsilon}^2)}\right )} \prod_{\sigma^4}{\prod_{\stackrel{\scriptstyle t-{\rm like}}{\sigma^2\subset\sigma^4}} \pr {{\rm d}^6 \tau_{\sigma^2|\sigma^4}}} \cdot \nonumber \\
\cdot \prod_{\sigma^4}{\prod_{\stackrel{{\scriptstyle\rm leaf/dia-}}{{\rm gonal~}\sigma^2}} \prback {{\rm d}^6 v_{\sigma^2|\sigma^4}}} \prod_{\sigma^3}{\D \Omega_{\sigma^3}}.
\end{eqnarray}

Projecting a path integral measure onto the physical hypersurface amounts to introducing some $\delta$-function factor. It is convenient to write it as the product of some two factors $\delta_{\rm metric}$ and $\delta_{\rm cont}$. The factor $\delta_{\rm metric}$ enforces the conditions
\begin{equation}                                                           
\epsilon_{abcd}v^{ab}_{\lambda\mu}v^{cd}_{\nu\rho}
\sim\epsilon_{\lambda\mu\nu\rho}
\end{equation}

\noindent ensuring that the area tensors correspond to certain edge vectors in each 4-simplex. These conditions are covariant in terms of the world indices. Therefore, the delta-function factor enforcing them is a scalar density. This agrees with the fact that the measures on the metric in the continuum GR are generally defined up to a power of $ - \det \| g_{\lambda \mu} \| $ \cite{Mis,DeW}. We can write \cite{Kha3} the general form of this factor which is a scalar density symmetrically as
\begin{equation}\label{delta(vv)}                                          
\delta_{\rm metric} ( v ) = \int V^\eta \delta^{21} \left ( \epsilon_{abcd}v^{ab}_{\lambda\mu}v^{cd}_{\nu\rho}
- V \epsilon_{\lambda\mu\nu\rho} \right ) \d V
\end{equation}

\noindent in the 4-simplex under consideration, and $\delta_{\rm metric}$ is the product of such expressions for the 4-simplices. The value of the parameter $\eta = 20$ is singled out by that then $\delta_{\rm metric}$ is a scalar, that is, it is invariant under an arbitrary deformation of the 4-simplex and thus can be considered to express itself some local property of the metric, not of the 4-simplex.

The factor $\delta_{\rm cont}$ ensures that the resulting edge lengths of any two neighboring 4-simplices coincide on their common 3-face. The situation when the two neighboring 4-simplices do not coincide on their common 3-face can be interpreted not necessarily as an ambiguity of the coordinates of the vertices of the common 3-face, but also as only a discontinuity on this 3-face of the metric induced from within each of these two 4-simplices. This allows to construct the $\delta$-function factor \cite{our5}; in particular, for the given 3-face it follows from the requirement of the invariance with respect to an arbitrary deformation of the 3-face leaving it in the same 3-plane. This $\delta$-function factor can be also found from the properly regularized formally infinite terms in the Einstein action in the path integral arising when substituting the discontinuous metric there \cite{our6}. Roughly speaking, in terms of area vectors, $\delta_{\rm cont}$ is the product of these factors over the 3-faces. Taking the area vectors $\bv_{\sigma^2_i }, i = 1, 2, 3$ of any three triangles $\sigma^2_i$ of $\sigma^3$ shared by 4-simplices $\sigma^4$ and $\sigma^{4 \prime}$, we can write this factor for $\sigma^3$ as
\begin{equation}\label{delta(vv-v'v')}                                     
\delta_{\rm cont}(\sigma^3) = [\bv_{\sigma^2_1 | \sigma^4 } \times \bv_{\sigma^2_2 | \sigma^4 } \cdot \bv_{\sigma^2_3 | \sigma^4 } ]^4 \delta^6 (\bv_{\sigma^2_i | \sigma^4 } \cdot \bv_{\sigma^2_j | \sigma^4 } - \bv_{\sigma^2_i | \sigma^{4 \prime} } \cdot \bv_{\sigma^2_j | \sigma^{4 \prime} }).
\end{equation}

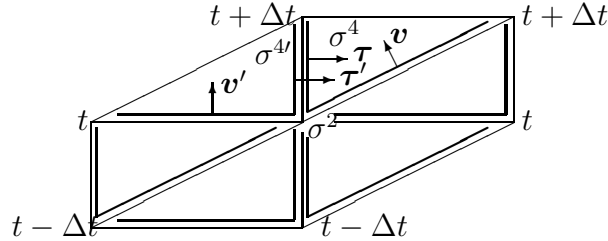
\begin{figure}[h]
\unitlength 1pt
\begin{picture}(150,95)(-138,10)
\put(40,60){\line(0,1){40}}
\put(40,60){\line(2,1){80}}
\put(40,100){\line(1,0){80}}
\put(40,100){\line(-2,-1){80}}
\put(40,60){\line(-1,0){80}}
\put(76,81){\vector(-1,2){5}}
\put(6,63){\vector(0,1){12}}
\put(73,90){$\bv$}
\put(9,71){$\bv^\prime$}
\put(42,84){\vector(1,0){15}}
\put(37,76){\vector(1,0){15}}
\put(54,74){$\btau^\prime$}
\put(59,83){$\btau$}
\put(50,89){$\sigma^4$}
\put(22,82){$\sigma^{4 \prime}$}
\put(42,54){$\sigma^2$}
\put(120,60){\line(0,1){40}}
\put(120,60){\line(-1,0){68}}
\put(40,20){\line(0,1){36}}
\put(40,20){\line(-1,0){80}}
\put(40,20){\line(2,1){80}}
\put(-40,20){\line(0,1){40}}
\put(-40,20){\line(2,1){80}}
\put(123,97){$t + \Delta t$}
\put(5,97){$t + \Delta t$}
\put(123,57){$t$}
\put(-45,57){$t$}
\put(-70,17){$t - \Delta t$}
\put(47,17){$t - \Delta t$}
\thicklines
\put(42,64){\line(0,1){34}}
\put(42,64){\line(2,1){68}}
\put(37,63){\line(0,1){34}}
\put(37,63){\line(-1,0){67}}
\put(117,63){\line(0,1){34}}
\put(117,63){\line(-1,0){65}}
\put(37,23){\line(0,1){34}}
\put(37,23){\line(-1,0){67}}
\put(42,24){\line(0,1){30}}
\put(42,24){\line(2,1){68}}
\put(-38,24){\line(0,1){34}}
\put(-38,24){\line(2,1){68}}
\end{picture}
\caption{4-simplices surrounding a given triangle $\sigma^2$. Fat lines in the 4-simplices show sets of triangles with tensors defined in these 4-simplices integration over which is present in the measure.}
\label{triangle}
\end{figure}

Important is that the factors $\delta_{\rm metric}$ and $\delta_{\rm cont}$ are of zero order in the scale of areas and, moreover, separately in the scale of the leaf/diagonal triangle tensors $\bv$ and the scale of the $t$-like triangle tensors $\btau$. Therefore, the order of the resulting measure in the scale of the leaf/diagonal triangle areas can be defined by power counting from (\ref{area-tensor-measure}) (at $ | \btau | \ll | \bv | $, since $\bv$ and $\btau$ enter the closure conditions $\sum \bv + \sum \btau = 0$ for the $t$-like 3-simplices). Namely, there are integrations in the measure (\ref{area-tensor-measure}) over twelve components of area tensors of each leaf/diagonal triangle $\sigma^2$ in the two containing it "future" 4-simplices $\sigma^4$ and $\sigma^{4 \prime}$, $\d^6 v_{\sigma^2 | \sigma^4 } \d^6 v_{\sigma^2 | \sigma^{4 \prime} }$, see Fig.\ref{triangle}.

Thus, if we try to approximate the area tensor part of the measure (\ref{area-tensor-measure}) after projecting onto the physical hypersurface by the product over separate areas and after separating out integrations over angle type variables, we can write out the factors $\rv^{11}_{\sigma^2} \d \rv_{\sigma^2}$ for the leaf/diagonal triangles $\sigma^2$. Together with the module of the result of integration over the connection part of the measure (\ref{N0}) (in the leading in $\varepsilon$ approximation) this gives the total measure in terms of the tetrad variables after integrating over $\D \Omega$. The resulting estimate for the measure referred to any leaf/diagonal triangle with the scale of the area tensor $\rv$ takes the form
\begin{equation}\label{Nv11dv}                                             
\N_0 \rv^{11} \d \rv = \left | \frac{1}{\frac{1}{4} \left ( \frac{1}{\gamma} - i \right )^2 \rv^2 + 1} \frac{\frac{1}{4} \left ( \frac{1}{\gamma} - i \right ) \rv}{ \sh \left [ \frac{\pi}{2} \left ( \frac{1}{\gamma} - i \right ) \rv \right ]} \right |^2 \rv^{11} \d \rv .
\end{equation}

\noindent Or $\N_0 \rv^{2 (\eta - 2) / 3 - 1} \d \rv $ if $\eta $ in (\ref{delta(vv)}) is not predefined as 20. In the physical spacelike region, $\rv^2 = - |\rv^2 |$.

Denoting the set of the scalar variables of the edge length type as $\rl = (l_1, \dots, l_n )$, we can write this total measure $F (\rl ) \d^n \rl $ in the sense of its dependence on the triangle areas in the factorization approximation as the product of the expressions of the type of (\ref{Nv11dv}) over the leaf/diagonal triangles,
\begin{equation}\label{F(l)dnl}                                            
F (\rl ) \d^n \rl \sim \prod_{\stackrel{{\scriptstyle\rm leaf/dia-}}{{\rm gonal~}\sigma^2}} \N_0 ( \rv_{\sigma^2} ) \rv_{\sigma^2}^{2 (\eta - 2) / 3 - 1} \d \rv_{\sigma^2}.
\end{equation}

\noindent Note that since the leaf/diagonal triangle areas present a redundant set of variables (in the case of a periodic 4-cube divided by diagonals into 24 4-simplices, there are 36 leaf/diagonal triangles per site and only 14 leaf/diagonal edges), the product of $\d \rv_{\sigma^2}$ over these triangles implies that there is a delta-function factor, invariant with respect to their overall scaling (at $\varepsilon << l$), which establishes relations between the areas $\rv_{\sigma^2}$.

Now consider the hypercubic structure. Analogously, the independent area tensor formulation can be introduced, in which $\bv_{\lambda\mu}$ in $S_{\rm cube} ( v, \Omega )$ are not expressions in terms of $l^a_\lambda$ (\ref{v=ll-R=wwww}), but are arbitrary complex vectors. This theory in the continuous time version has the commuting constraints and a simple canonical measure of the type (\ref{dvDOmega}). Going back to the full discrete theory, we have formally the same expression for the measure on the independent area tensors (\ref{area-tensor-measure}) with the corresponding replacement of $\sigma^n$ by the tensor notation,
\begin{equation}                                                           
\prod_{\rm sites} \left \{ \left [ \delta (\btau_1 \cdot \btau_2) \prod^3_{\alpha = 1} \delta (\btau^2_\alpha - \tilde{\varepsilon}^2) \right ] \left ( \prod^3_{\alpha = 1} \d^6 \tau_\alpha \right ) \left ( \prod^3_{\alpha = 1} \d^6 v_\alpha \right ) \left ( \prod^3_{\lambda = 0} \D \Omega_\lambda \right ) \right \}
\end{equation}

\noindent and the same $\delta$-function factor $\delta_{\rm metric} ( v )$ (\ref{delta(vv)}). Here it is taken into account that $\tau_\alpha \circ \tau_\beta = \btau_\alpha \cdot \btau_\beta$ modulo $\delta_{\rm metric} ( v )$ (by the way, the same substitution can be made in (\ref{area-tensor-measure})). We have a simplification in the form of a simpler combinatorics, the absence of continuity conditions on the 3-faces (the factor $\delta_{\rm cont}$ (\ref{delta(vv-v'v')})) and of redundant variables. The measure is defined in a closed form as a product over the 4-cubes or sites of the following factors,
\begin{equation}\label{int-d36v-delta-metric}                              
\int \d^{36} v^{a b}_{\lambda \mu} \delta_{\rm metric} ( v ) \sim [\bv_1 \times \bv_2 \cdot \bv_3 ]^{( \eta - 8 )/ 2} (l^0_0 )^{(\eta - 6)} \d^9 \bv_\alpha \d^4 l_0 {\cal DO}.
\end{equation}

\noindent Here ${\cal O} \in \mbox{O(3, C)}$ is inversion plus rotation by an imaginary angle which links the anti-selfdual triad with the selfdual one, $\bv^*_\alpha = {\cal O} \bv_\alpha$; this freedom is fixed by using time gauge $l^0_\alpha = 0, \alpha = 1, 2, 3$ (then ${\cal O} = -1$), which is just used in the rest part of the measure. In the underlying theory of independent area tensors, we fix some degrees of freedom (four per site) of the area tensors $\tau_\alpha$, but not of $l^a_0$. Then the scale of $l^a_0$ is $|\btau | / \sqrt{\rv}$ where $|\btau |$ is the scale of the area vectors $\btau$. This dependence on $\rv$ violates the factorization over separate $\rv_\alpha$. In the factorization approximation, we distribute the dependence on $\rv$ multiplicatively equally as dependencies on different $\rv_\alpha$. The resulting area tensor part of the measure together with the module of the result of integration over the connection part of the measure (\ref{N0}) takes the form
\begin{equation}\label{Fcube(l)dnl}                                        
F_{\rm cube} (\rl ) \d^n \rl \sim \prod_{\rm sites} \left [ [\bn_1 \times \bn_2 \cdot \bn_3 ]^{( \eta - 8 )/ 2} \d^6 \bn \right. \prod_\alpha \left. \N_0 ( 2 \rv_{\alpha} ) \rv_{\alpha}^{ (\eta - 2) / 3 - 1} \d \rv_{\alpha} \right ],
\end{equation}

\noindent where $\bn_\alpha = \bv_\alpha / \rv_\alpha$, $\d^6 \bn \equiv \d^2 \bn_1 \d^2 \bn_2 \d^2 \bn_3$. Here we have written also the angle part of the measure originating from a power of $[\bv_1 \times \bv_2 \cdot \bv_3 ]$. Its meaning is that it suppresses configurations that are close to degenerate ones with $[\bv_1 \times \bv_2 \cdot \bv_3 ] = 0$. The matter is that we have only three areas $\rv_\alpha$ per 4-cube which could be fixed by maximizing the measure; at the same time, maximizing the measure also means maximizing $[\bn_1 \times \bn_2 \cdot \bn_3 ]$, and this gives an orthogonal triple as an optimal choice for $\bv_1, \bv_2, \bv_3$ and thus for $\bl_1, \bl_2, \bl_3$.

Of course, such factors, powers of $[\bv_{\sigma^2_1} \times \bv_{\sigma^2_2} \cdot \bv_{\sigma^2_3} ]$, are also present in the simplicial measure $F (\rl ) \d^n \rl$, since we have the same formula (\ref{int-d36v-delta-metric}) in each 4-simplex, but the number of the leaf/diagonal triangle areas which are fixed by maximizing the measure is now sufficient to fixe the leaf/diagonal edge lengths. Therefore, these determinant-type factors are not shown in the estimate for $F (\rl ) \d^n \rl$ (\ref{F(l)dnl}) as compared with the dependence on $\rv$, which has more pronounced maxima for large $\eta$. Nevertheless, these factors may also be relevant in the simplicial case in order to suppress accidental configurations with "spikes", when triangles with fixed areas can have arbitrarily large sides.

The dependence on area scales in (\ref{F(l)dnl}) and (\ref{Fcube(l)dnl}) differs due to combinatorial-topologi\-cal differences and can be written uniformly as a product of the factors
\begin{equation}                                                           
\rv^{(\eta - 2) N_4 / L_2 - 1} \d \rv
\end{equation}

\noindent over areas. Here $L_2$ is the number of the leaf/diagonal triangles in the simplicial case or the quadrangles of the 3D space sections in the hypercubic case; $N_4$ is the number of the 4D elementary regions (4-simplices or 4-cubes). That is, $L_2 / N_4$ is the number (per elementary 4D-region) of the areas described by the tensor components which are analogs of space-like ones over {\it world} vector indices. There are four $\sigma^4$ in each elementary 4-prism and three leaf/diagonal $\sigma^2$ in each elementary 3-prism between any two neighboring leaves. Let $N^{(3)}_k$ be the number of the $k$-simplices in a 3D leaf; then in the simplicial case
\begin{equation}                                                           
N_4 / L_2 = 4 N^{(3)}_3 / (3 N^{(3)}_2 ) = 2/3,
\end{equation}

\noindent since $N^{(3)}_2 = 2 N^{(3)}_3$. In the hypercubic case, $N_4 / L_2 = 1/3$.

\section{Condition on the background lengths for the perturbation expansion and background length scale estimate}\label{perturbation}

Here we define background lengths as an optimal starting point of the perturbation series in the obtained purely tetrad/length theory. It is important for this definition through the maximum of (\ref{def-l0}) that $\det \| \partial^2 S / \partial l_i \partial l_k \| \neq 0$, and this point is briefly discussed. Using the factorization approximation for the tetrad part of the measure, we estimate the background area scale $\rv_0$ which maximizes (\ref{def-l0}) or, in terms of it, the length scale $a$,
\begin{equation}                                                           
a = \sqrt{2 | \rv_0 |},
\end{equation}

\noindent for the general simplicial (actually $a$) (\ref{a=sqrt}) and hypercubic system $a_{\rm cube}$ (\ref{acube=sqrt}). Introducing $a$ is somewhat conventional, but its hypercubic version $a_{\rm cube}$ has a more exact sense of the 3-cube edge length, since the optimal choice of the background triad $\bl_1, \bl_2, \bl_3$, maximizing the measure, means its orthogonality, as discussed below (\ref{Fcube(l)dnl}), and the triangles, whose area is estimated to be $| \rv_0 | = a^2 / 2$, are rectangular.

We have the result of the functional integration over connection of the form
\begin{equation}\label{FexpiS(l)}                                          
\int F (\rl ) \d^n \rl \exp (i S (\rl ) ).
\end{equation}

\noindent Here $\rl = (l_1, \dots, l_n )$ is the set of the edge lengths, $S(\rl )$ is the Regge action, the measure $F (\rl ) \d^n \rl $ in the sense of its dependence on the triangle areas was approximately given as products over the leaf/diagonal triangles, (\ref{F(l)dnl}) and (\ref{Fcube(l)dnl}) in the hypercubic version. $F (\rl )$ has maxima at finite areas and rapidly decreases if any area increases (exponentially).

On the flat background, the Regge equations of motion are satisfied identically for any edge lengths. Then the system is governed by $F (\rl )$ to be in the vicinity of its maxima. For an arbitrarily small curvature, these equations abruptly change the solution and seem to control the lengths; however, in the functional integral, there is no reason for arising such an abruptness. It is natural to expect that the effect is defined by the angle defects $\alpha_{\sigma^2}$. With a typical curvature $R$, a typical defect
\begin{equation}\label{alpha-ll-1}                                         
\alpha \sim R a^2, \mbox{~ and ~} \alpha << 1
\end{equation}

\noindent for $R$ encountered in practice and $a \sim 1$; then the system is governed by $F (\rl )$.

Consider constructing a formal perturbation theory. We pass from $\rl$ to a new $n$-vector variable $u = (u_1, \dots, u_n )$ which makes the measure to be the Lebesgue one. We make the Taylor expansion of $S(\rl )$ around some point $\rl_0 = \rl ( u_0 )$ ($u_0 = (u_{01}, \dots, u_{0n} )$, $\rl_0 = (l_{01}, \dots, l_{0n} )$) over $\Delta u = u - u_0$,
\begin{eqnarray}                                                           
F ( \rl ) \d^n \rl = \d^n u, ~~~ S ( \rl ( u ) ) = S ( \rl_0 ) + \sum_{j, k} \frac{ \partial S ( \rl_0 )}{ \partial l_j } \frac{ \partial l_j ( u_0 )}{ \partial u_k } \Delta u_k \nonumber \\ + \left. \frac{1}{2} \sum_{j, k, l} \frac{\partial }{\partial u_l } \left ( \frac{ \partial l_j }{ \partial u_k } \frac{ \partial S }{ \partial l_j } \right ) \right |_{u = u_0} \Delta u_k \Delta u_l + \dots .
\end{eqnarray}

\noindent The requirement that there be no term linear in $\Delta u$ in the latter means the classical equations of motion (the Regge equations),
\begin{equation}\label{Sdudu}                                              
\frac{\partial S (\rl_0 )}{\partial \rl} = 0, ~~~ S (\rl ) = \frac{1}{2} \sum_{j, k, l, m} \frac{\partial^2 S (\rl_0 )}{\partial l_j \partial l_l} \frac{\partial l_j (u_0 )}{\partial u_k} \frac{\partial l_l (u_0 )}{\partial u_m} \Delta u_k \Delta  u_m + \dots .
\end{equation}

Is there any reason that $\det \| \partial^2 S (\rl_0 ) / \partial l_i \partial l_k \| $ be identically zero? Consider the flat background spacetime. Any skeleton $(l_1, \dots, l_n )$ that is realizable in this spacetime satisfies the Regge equations. Therefore, the matrix of the second variation of $S$ for the variations of the lengths leaving the skeleton in the flat spacetime is identically zero. Let us compare the dimensionality of this matrix and of the full one for arbitrary variations of the lengths. One can imagine the 4-dimensional spacetime as a hypersurface embedded in a flat spacetime of a sufficiently large dimensionality \cite{RegTeit} $D$ with the pseudoEuclidean metric $\eta_{AB} = {\rm diag} ( -1, +1, +1, \dots, +1 ) $. Let the flat 4-dimensional spacetime be the set of points with the coordinates $x^A$ different from zero only at $A = 0, 1, 2, 3$ (a hyperplane). The coordinates of the vertices $x^A_{\sigma^0}$ for any edge $\sigma^1$ with the ending vertices $\sigma^0_1, \sigma^0_2$ define its vector $l^A_{\sigma^1} = x^A_{\sigma^0_1} - x^A_{\sigma^0_2}$ and, in particular, the length $l_{\sigma^1}$. The motion of a vertex by $\delta x^A_{\sigma^0}$ can be decomposed into its translation in the hyperplane $\delta_{\| } x_{\sigma^0} = (*, *, *, *, 0, \dots, 0)$ ($\delta_{\| } x^A_{\sigma^0} = 0$ at $A \geq 4$) and physical fluctuation $\delta_{\bot} x_{\sigma^0} = (0, 0, 0, 0, *, *, \dots, *)$ ($\delta_{\bot } x^A_{\sigma^0} = 0$ at $A = 0, 1, 2, 3$), which makes the flat spacetime curved. The translations $\delta_{\| } x_{\sigma^0}$ or $\delta_{\| } l_{\sigma^1}$ leave the spacetime flat and generate gauge transformations on it \cite{RocWilPL,RocWil,Wil}. In particular, $\delta_{\| }^2 S = 0$ on the flat spacetime. At the same time, $\delta_{\| } \delta_{\bot} S$ is generally nonzero even on the flat spacetime (since $\delta_{\| } S$ is nonzero on the curved spacetime), as well as $\delta_{\bot }^2 S$. In overall, (the matrix of) the quadratic form $\delta^2 S$ on the flat spacetime where $\delta$ is $\delta_{\| }$ or $\delta_{\bot }$ has zero block $\delta_{\| }^2 S$, but its dimension (4 parameters per vertex) is less than half of the dimension of $\delta^2 S$ itself ($D \geq 10$ parameters per vertex). Therefore $\det \| \partial^2 S / \partial l_i \partial l_k \| \neq 0$ for a random skeleton (with proper initial/boundary conditions imposed). This is already without the procedure of fixing some edge lengths, similar to the continuum gauge fixing.

If the integral $\int \d^n u = \int F (\rl ) \d^n \rl$ is finite (which is indeed the case), the boundary of the range of the variable $u$ is located mainly at finite $u$; when approaching this boundary, then $D(l_1, \dots , l_n)/ D(u_1, \dots , u_n) = 1/ F (\rl ) \to \infty$. That is, this shows up as an infinite potential wall in $S( \rl (u ) )$. Since, as it turns out, $F (\rl )$ is zero if any area is zero or infinity, the system does not go to these points, but on the contrary, the dominant contribution to the path integral comes from the neighborhood of the point $\rl = \rl_0$ which minimizes the determinant of the considered form $\delta^2 S$ in the action (\ref{Sdudu}) or maximizes the inverse of it,
\begin{equation}\label{def-l0}                                             
F (\rl_0 )^2 \det \left \| \frac{\partial^2 S (\rl_0 )}{\partial l_i \partial l_k} \right \|^{-1}.
\end{equation}

Although, an accidental additional symmetry of the skeleton is also possible leading to $\det \| \partial^2 S (\rl_0 ) / \partial l_i \partial l_k \| = 0$ \cite{RocWilPL,RocWil,Wil}; let the rank of the $n \times n$ dimensional matrix $ \partial^2 S / \partial l_i \partial l_k $ be, say, $n - 1$. We can consider (combinations of) the lengths $(l_1, \dots, l_n )$ which are chosen so that $\delta^2 S$ at $\rl = \rl_0$ does not depend on $\delta l_n$. We can fix the gauge by fixing $l_n$ \cite{RocWilPL} and omitting the integration over $\d l_n$. As a result, we also have (\ref{def-l0}), where now $\partial^2 S / \partial l_i \partial l_k$ is a $(n-1) \times (n-1)$ dimensional matrix: $i, k = 1,2, ..., n-1$.

Thus, in order to find the initial point of the perturbative expansion, we, as usual, solve the equations of motion. Here we take the flat spacetime as such a solution. The variables which are not defined by the equations of motion, the edge lengths of the (flat) skeleton, are defined by maximizing (\ref{def-l0}).

Analogously, the equations of motion for $S_{\rm cube} (\rl )$ do not define the length scale, which is defined by maximizing (\ref{def-l0}).

Important is invariance of the definition of the physical point in the configuration superspace at $\rl = \rl_0$ by maximizing (\ref{def-l0}) if we use new variables $q$ related to $\rl$ by a nondegenerate redefinition $\rl \to q = q ( \rl )$.

A simple estimate can be made if the dependence on the scale of areas of a group of triangles is singled out.
The product of the measures of the type (\ref{Nv11dv}) over some number $T$ of the triangles in a certain domain contains a measure on their common area scale $\rv$ (expectedly, $\rv$ is close to the maxima of separate expressions of the form (\ref{Nv11dv})) of the type $(\N_0 \rv^{12})^T \rv^{-1} \d \rv$. Since the action depends linearly on $\rv$, $\partial^2 S / \partial \rv_i \partial \rv_k$ has the scale $1/ \rv$ and the optimal point $\rv = \rv_0$ is defined by the maximum of $(\N_0 \rv^{12})^T \rv^{-1 / 2} $ or, at large $T$, the maximum of $\N_0 \rv^{12}$. Or $\N_0 \rv^{2 (\eta - 2) / 3}$ if $\eta $ in (\ref{delta(vv)}) is not predefined. Here we considered dependence on the triangle area scale $\rv$, but could act in terms of the length scale $l$, $2 \rv = i l^2$, which would set the task of maximizing the same function. There are two competing mechanisms for arising a maximum of this function.

1) Closeness to a pole (in the unphysical region $\Im \rv^2 \neq 0$). At $\gamma \ll 1$, there is a local maximum at $\rv^2 = -4 \gamma^2$ (and smaller maxima at $\rv^2 = -4 \gamma^2 n^2, n \ll \gamma^{-1}$).

2) Interplay between a power function and a decreasing exponent. At $ | \rv | \gg 1$, the function behaves like $|\rv |^{2 (\eta - 5) / 3} \exp {(- \pi |\rv |)}$ and has a maximum at $|\rv | = |\rv_0 | = 2 (\eta - 5) / (3 \pi ) $. (Thus, the largeness of $\rv$ assumes that $\eta$ is large, and $\eta = 20$, selected after (\ref{delta(vv)}), can be a boundary value in this sense.)

An estimate for the ratio of the maximum (ii) to the maximum (i), $( 4 \pi )^2 [(\eta - 5 )(3 e \pi )^{-1}]^{2 (\eta - 5 )/3 } \gamma^{- 2 (\eta - 11 ) / 3}$, is $>1.5$ orders for the selected $\eta = 20$ and $\gamma = 1/2$ as some upper bound for "small" $\gamma $. The maximum (ii) at large $\eta$ dominates.

Then the above used length scale $a$, $2 \rv_0 = i a^2$, can be estimated,
\begin{equation}\label{a=sqrt}                                             
a = \sqrt{ 4 ( \eta - 5) / (3 \pi ) }, \mbox{~~~ in usual units ~} a = \sqrt{ 32 G ( \eta - 5) / 3 },
\end{equation}

\noindent where $\eta = 20$ ($a = 4 \sqrt{ 10 G }$) can be singled out, as mentioned after (\ref{delta(vv)}). For the hypercubic structure, we have half the degree of dependence of the tetrad part of the measure on the area, $F_{\rm cube}$ (\ref{Fcube(l)dnl}) vs $F$ (\ref{F(l)dnl}), and, besides, $\N_0 (2 \rv )$ in $F_{\rm cube}$ vs $\N_0 (\rv )$ in $F$. This combinatorial-topological difference leads to
\begin{equation}\label{acube=sqrt}                                         
a_{\rm cube} = \sqrt{ ( \eta - 8) / (3 \pi ) }, \mbox{~~~ in usual units ~} a_{\rm cube} = \sqrt{ 8 G ( \eta - 8) / 3 }
\end{equation}

\noindent ($a_{\rm cube} = 4 \sqrt{ 2 G }$ for $\eta = 20$).

The background length scale is directly related to the parameter $\eta$, which parameterizes $\delta_{\rm metric} ( v )$ (\ref{delta(vv)}). Varying $\eta$ by $\Delta \eta$ results in additional 4D volume factors $V^{\Delta \eta}_{\sigma^4}$ for the 4-simplices $\sigma^4$ in the measure. These are just the factors which correspond to the factors $ ( - \det \| g_{\lambda \mu} \| )^{\Delta \eta / 2}$ in the continuum GR. There is a reference value $\eta = 20$ for which $\delta_{\rm metric} ( v )$ is a scalar with respect to the world indices, but if we assume that $\delta_{\rm metric} ( v )$ is a general scalar density, then $\eta$ is a free constant of the theory (which does not manifest itself at the classical level). This match the assumption that the continuum measure is not a scalar, but a general scalar density.

The above estimates (\ref{a=sqrt}) and (\ref{acube=sqrt}) show how the dependence on $\eta$ is sensitive to the topological differences between the models. However, if $\eta$ is taken as a free constant, it (or its effective value) can be different in each theory, say, $\eta$ and $\eta_{\rm cube}$.

Too small $\eta$ leads to a formal divergence of the functional integral at the lower limit. Then this integral is saturated by arbitrarily small lengths, and we have $a = 0$, so the theory becomes continuous {\it dynamically} and the discrete regularization is taken off. It looks like a kind of phase transition to the continuum phase. However, before we reach $a = 0$, the estimates (\ref{a=sqrt}) and (\ref{acube=sqrt}) should be modified, since these imply large resulting values, but more importantly, the stationary phase expansion for the phase blows up, and we can not confine ourselves to the GR action term in it as an actual action.

The dependence on $\eta$ in the continuum analogue of our integration over connection and projecting the measure from the superspace of independent area tensors $v^{a b}_{\lambda \mu}$ to the superspace of the tetrad/metric with the help of $\delta_{\rm metric} ( v )$ looks as
\begin{eqnarray}\label{eta-continuum}                                      
& & \int \exp [iS(e, \omega)] \prod_x [\d^{24} \omega^{a b}_\lambda \d^{36} v^{a b}_{\lambda \mu} \delta_{\rm metric} ( v )] \nonumber \\ & & = \int \exp [iS(e, \omega (e ))] \prod_x [(- \det \| g_{\lambda \mu} \|)^{(\eta - 13)/ 2} \d^{10} g_{\lambda \mu}].
\end{eqnarray}

\noindent (This integration goes through powers of $\det \| e^a_\lambda \| = ( - \det \| g_{\lambda \mu} \| )^{1 / 2}$ which can be easily found from the dimension considerations, with taking into account the dimension $[\omega ] = [ e ]^{- 1}$.) It is seen that the continuum counterpart $(- \det \| g_{\lambda \mu} \|)^\alpha \d^{10} g_{\lambda \mu}$ of our measure (if we take the reference value $\eta = 20$ so that $\delta_{\rm metric}$ is a scalar) possesses rather large positive power $\alpha = 7 / 2$ as compared to the Misner \cite{Mis} ($\alpha= - 5 / 2$) or DeWitt  \cite{DeW} ($\alpha = 0$) measures.

Note that the possibility of suppressing small lengths/areas due to the standard phase volume type tetrad measure at not too small $\eta$ has appeared due to the exponential suppression of large areas contained in the module of the result of the functional integration over connection of the type $\N_0 (\rv )$ (\ref{N0}). Without such a suppression, integrals over these power type measures would be divergent at the upper limit, and would not have such a simple interpretation.

\section{On the form of the propagator}\label{propagator}

We consider some general features of the propagator, paying particular attention to the case of $S_{\rm cube} (v, \Omega )$, where $S_{\rm cube} (v, \Omega_0 ( v ) )$ ($\Omega_0 (v )$ is a particular solution of the equations of motion for $\Omega$) is a finite-difference form of the GR action in the leading approximation for a slowly varying metric from site to site (expansion over these variations $\delta$ can be defined in a regular way), for which the propagator can be given in a closed form.

In accordance with the considered mechanism for fixing field variables (the leaf/diago\-nal edge lengths are loosely fixed dynamically while the discrete lapse-shift vectors are given as parameters) the propagator of interest is the simplicial analog of the continuum GR propagator in the synchronous frame, $\d s^2 = \gamma_{\alpha \beta} \d x^\alpha \d x^\beta - \d t^2$ (this corresponds to the lapse-shift $l^0_0 = 1, \bl_0 = 0$), for which the continuum action has the form
\begin{eqnarray}\label{int-R-sqrt-g-d4x}                                   
\frac{1}{2} \int R \sqrt{-g} \d^4 x = \frac{1}{4} \int \d^4 x \sqrt{-g} \left \{ \frac{1}{2} \left ( \gamma^{\alpha \gamma} \gamma^{\beta \delta} - \gamma^{\alpha \beta} \gamma^{\gamma \delta} \right ) \dot{\gamma }_{\alpha \beta} \dot{\gamma }_{\gamma \delta} \right. \nonumber \\
\left. + \left [ \gamma^{\alpha \zeta} \left ( \gamma^{\beta \delta} \gamma^{\epsilon \gamma} - \gamma^{\beta \gamma} \gamma^{\epsilon \delta} \right ) + \frac{1}{2} \gamma^{\gamma \zeta} \left ( \gamma^{\alpha \beta} \gamma^{\epsilon \delta} - \gamma^{\alpha \epsilon} \gamma^{\beta \delta} \right ) \right ] \gamma_{\alpha \beta , \gamma } \gamma_{\delta \epsilon , \zeta } \right \},
\end{eqnarray}

\noindent and the propagator for the perturbations $\Delta \gamma _{\alpha \beta}$ of the background metric $\gamma _{\alpha \beta}$ in accordance with the standard definition using a source term in the action,
\begin{eqnarray}                                                           
S_J = \frac{1}{2} \int R \sqrt{-g} \d^4 x + \int J^{\alpha \beta} \Delta \gamma _{\alpha \beta} \sqrt{-g} \d^4 x, \nonumber \\
\Delta \gamma _{\alpha \beta} (x ) = - \int G_{\alpha \beta}{}^{\gamma \delta} (x, y) J_{\gamma \delta} (y ) \sqrt{-g} \d^4 y,
\end{eqnarray}

\noindent takes the form in the momentum representation (if the background $g_{\lambda \mu} = const$)
\begin{eqnarray}\label{G-full}                                             
G^{\gamma \delta}{}_{\alpha \beta} \sqrt{- g } = 2 \frac{ L^\gamma_\alpha L^\delta_\beta + L^\delta_\alpha L^\gamma_\beta - L^{\gamma \delta} L_{\alpha \beta} } {p^2_0 - \bp^2}, ~~~ L_{\alpha \beta} = \gamma_{\alpha \beta} - \frac{p_\alpha p_\beta}{p^2_0} .
\end{eqnarray}

\noindent Here $\bp^2 = p_\alpha p_\beta \gamma^{\alpha \beta}$. At $l^0_0 \neq 1, \bl_0 = 0$, the replacement $p_0 \to p_0 (l^0_0 )^{-1}$ should be made. We also note that there are no Faddeev-Popov ghosts in the continuum GR for this gauge, therefore the analogy with the discrete case, in which the corresponding gauge symmetries are absent, turns out to be quite accurate.

In the simplicial case, the background lengths and the propagator are functionals of the simplicial structure, that is, of the coincidence matrix showing which vertices are connected by edges. If we compute any diagrams on the basis of this propagator, the result should be averaged over the possible simplicial structures. Thus far it seems to be not an easy task.

Simplification occurs when the set of variables is reduced, while still remaining sufficient in the continuum limit, that is, in the hypercubic model.

Above we consider the implicitly introduced variables $u$ reducing the measure to the Lebesgue one (in Section \ref{perturbation}), but to write out the propagator it is more convenient to use the original length or metric variables. This means that instead of $\langle \Delta u_j \Delta u_k \rangle$ we consider this correlator, linearly transformed to the variables of the squared lengths $s_j = l^2_j$ or metric, $\sum_{l, m}( \partial s_j ( u_0 ) / \partial u_l ) \langle \Delta u_l \Delta u_m \rangle \partial s_k ( u_0 ) / \partial u_m$.

In the hypercubic model, the optimal choice for $\bv_\alpha$ implies an orthogonal triple $\bl_1, \bl_2, \bl_3$, as discussed below (\ref{Fcube(l)dnl}), and the optimal values of three lengths $| \bl_\alpha |$ being $a_{\rm cube}$ (\ref{acube=sqrt}). This means the background metric $\gamma_{\alpha \beta} = a_{\rm cube}^2 \delta_{\alpha \beta}$.

As for the time metric components, temporarily discussing also the more general simplicial case, we fixe the scale of the $t$-like area tensors at the level $\tilde{\varepsilon}$ (\ref{vv=e2}) and thus fixe $l^a_0$: $\varepsilon = l^0_0 = 2 \tilde{\varepsilon} a^{-1}, \bl_0 = 0$. There are two reference values of $l^0_0$: an arbitrarily small value or $a$.

The choice $\varepsilon = a$ instead of $\varepsilon \ll a$ improves the accuracy of the approximation of the effective action (which can be thought to be the main definitive factor for the form of the propagator) by the leading term of the stationary phase expansion, since this diminishes the parameter of this expansion $(\varepsilon a)^{-1}$; the accuracy of the expansion of the module of the result of the functional integration becomes worse, but one can expect that this module in the given calculation defines mainly the background length scale $a$, but not the form of the propagator. Thus, it may be expedient to illustrate the form of the propagator by approaching from the relatively small values of $\tilde{\varepsilon}$ to the symmetric version $\tilde{\varepsilon} = a^2 / 2, \varepsilon = l^0_0 = a$.

In the hypercubic model, this means the background metric $g_{\lambda \mu} = a_{\rm cube}^2 \eta_{\lambda \mu}$, $\eta_{\lambda \mu} ={\rm diag}(-1,1,1,1)$.

If we start from the small variations $\delta_\lambda$ of the tetrad/metric, $S_{\rm cube} (v, \Omega)$ in the leading approximation is a finite difference form of the Cartan-Weyl-Holst action (\ref{Cartan}), excluding $\omega$ ($\Omega = \exp \omega$) from which gives a finite difference form of $\frac{1}{2} \int R \sqrt{-g} \d^4 x$ or, in the considered gauge, of (\ref{int-R-sqrt-g-d4x}).

It is convenient to express $\delta_\lambda$ in terms of the symmetric form $\delta^{\rm sym}_\lambda = ( T_\lambda - \overline{T}_\lambda ) / 2 = ( \delta_\lambda - \overline{\delta}_\lambda ) / 2 = \delta_\lambda - \delta_\lambda \overline{\delta}_\lambda / 2$ for $\partial_\lambda$; in the leading approximation in $\delta$, $\delta_\lambda$ is substituted by $\delta^{\rm sym}_\lambda$. In the momentum representation, $T_\lambda = \exp (i p_\lambda )$, $\delta^{\rm sym}_\lambda = i \sin p_\lambda$, where $p_\lambda$ is now a quasimomentum. This definition has the property $\overline{\delta}^{\rm sym}_\lambda = - \delta^{\rm sym}_\lambda$, the same as $\overline{\partial}_\lambda = - \partial_\lambda$, which allows transferring $\delta^{\rm sym}_\lambda$ from one factor to the other factors when integrating the quadratic form in $ \int R \sqrt{-g} \d^4 x$ by parts, and the discrete propagator follows simply by the replacement of $ p_\lambda$ by $ - i \delta^{\rm sym}_\lambda = \sin p_\lambda$ in the continuum version.

For clarity, it is useful for the propagator to write the following value:
\begin{eqnarray}\label{G-full-discr}                                       
& & \hspace{-7mm} a_{\rm cube}^2 G_{\!\! \rm cube}^{~\,\gamma \delta}{}_{\alpha \beta} = 16 \pi G \frac{\cal L^\gamma_\alpha L^\delta_\beta + L^\delta_\alpha L^\gamma_\beta - L_{\alpha \beta} L^{\gamma \delta} }{\sin^2 \! \! p_0 - \sum_\alpha \sin^2 \! \! p_\alpha}, ~~~ {\cal L_{\alpha \beta}} = \gamma_{\alpha \beta} - \frac{\sin p_\alpha \sin p_\beta}{\sin^2 \! \! p_0}
\end{eqnarray}

\noindent (in the usual units, with the action $(16 \pi G)^{-1} \int R \sqrt{-g} \d^4 x$). Since $a_{\rm cube}^3 \gamma^{\gamma \alpha} \gamma^{\delta \beta} \Delta \gamma_{\alpha \beta}$ and $a_{\rm cube}^{-1} \Delta \gamma_{\alpha \beta}$ have the meaning of length variations $\Delta l_j$, the LHS is a correlator of the type of $\langle \Delta l_j \Delta l_k \rangle$ (with $\Delta l_j \approx \sum_k ( \partial l_j ( u_0 ) / \partial u_k ) \Delta u_k $).

For non-small $p_\lambda$, higher order corrections in $\delta$ to the quadratic form of the action can be essential. They are obtained by expanding over $\delta$ and $\omega$ the curvature form (\ref{r}), and then $S_{\rm cube} (v, \Omega)$ (\ref{S-cube}), which depends on this form, solving the equations of motion for $\omega$ by the method of successive approximations to the leading approximation $\omega = O( \delta )$ and substituting $\omega$ back into $S_{\rm cube}$.

An interesting question arises if we would like to approximate the diagrams of the more general simplicial theory by the diagrams of such a theory with the reduced set of variables, the hypercubic one. Our still free parameter $\eta$ can be different in each theory, the actual $\eta$ in the general simplicial theory and $\eta_{\rm cube}$ in the hypercubic theory. The hypothesis is that the parameters $a$ and $a_{\rm cube}$, which have a more direct physical meaning, should correspond to each other for our approximation goal,
\begin{equation}                                                           
a_{\rm cube} ( \eta_{\rm cube} ) = a ( \eta ),
\end{equation}

\noindent that is, $a$ could be used in $G_{\!\! \rm cube}^{~\,\gamma \delta}{}_{\alpha \beta}$ for a rough approximation. Especially because in another respect there is already the mentioned above correspondence near the continuum limit (via the relations between the variables of the type (\ref{omega-lambda=sum-omega-sigma})) between $S ( v, \Omega )$ for the simplest periodic Regge manifold and $S_{\rm cube} ( v, \Omega )$ on the level of the quadratic connection forms.

\section{Conclusions}

We consider the procedure of constructing the perturbative expansion with taking into account the mechanism of arising length scales non-perturbatively.

Upon functional integration of $\exp (iS(v, \Omega))$ over $\Omega$, we are left with the functional integral over the tetrad type variables of the expression $F \exp (iS)$.

In $F \exp (iS)$, we can take for $F$ the leading order term of the expansion over the scale $\varepsilon$ of the discrete lapse-shift functions and for $S$ the leading order term of the stationary phase expansion $S(v, \Omega_0 (v ))$ ($\Omega_0 (v )$ is a particular solution of the equations of motion for $\Omega$), that is, Regge action. We can confine ourselves to the leading terms of these expansions in some region of the parameters $\varepsilon$, $a$ of the type (\ref{a-1-e-a}). Here $a$ is the length scale of the leaf/diagonal edges defined inside our approach by the constants of the theory $ \eta , \gamma $ where $\eta$ defines a discrete analogue of the volume factor in the measure like $(  - \det \| g_{\lambda \mu} \| )^{\eta / 2}$.

In parallel with the general Regge manifold, the connection action $S (v, \Omega)$ is also considered on the simplest periodic Regge manifold, and some part of (combinations of) the variables $v, \Omega$ is frozen. Periodicity means a hypercubic cell. For such an action, $S_{\rm cube} (v, \Omega)$ (\ref{S-cube}), for the result of the functional integration over connection, we can take $F_{\rm cube} \exp (iS_{\rm cube}  (v, \Omega_0 ( v )))$, where $S_{\rm cube}  (v, \Omega_0 ( v ))$ is an action, which tends to the GR action for a slowly varying metric from site to site.

The perturbative expansion is around a point which is fixed not only by re\-qu\-ir\-ing extremum of the zero-order term (via equations of motion) in Taylor ex\-pan\-sion of action, but also by minimizing determinant of the second order form (in variables in which the measure is Lebesgue) or by maximizing (\ref{def-l0}) or, roughly, the measure and thus fixing, in particular, the length scale. In Section \ref{perturbation} we dis\-cuss this for a flat background spacetime and note approximate validity of this procedure also at small defects (\ref{alpha-ll-1}) (in particular, for curvatures usually encountered in practice); the length scale $a$ is estimated for the general simplicial system (\ref{a=sqrt}) and for the reduced (hypercubic) one (\ref{acube=sqrt}), the difference in the dependence on the free parameter $\eta$ is of a combinatorial-topological origin.

The discrete propagator is a simplicial analog of the continuum graviton propagator in {\it the synchronous frame gauge} (or generalized synchronous frame gauge with arbitrary lapse-shift). In principle, we should know it as a functional of the background Regge lattice appropriately fixed at the length scale estimated in order to average the result of any diagram calculations over the lattice structures. A closed form available for calculations is for the system described by $S_{\rm cube} (v, \Omega )$ in the leading approximation for small variations $\delta$ of the metric from site to site (\ref{G-full-discr}) (subsequent orders in $\delta$ can be defined in a regular way).

The above can be compared to the continuum theory in that, roughly speaking, the diagrams originally divergent as a power of a momentum cut off $\Lambda$ are now finite and proportional to the same power of $a^{-1}$, and we have an expansion in powers of $a^{-2}$.

A peculiarity of the theory is the implicit use of the variables $u$, in which the measure is Lebesgue. In particular, the propagator under consideration, as already mentioned, is in fact a correlator for $\Delta u$, linearly transformed to $\langle \Delta l_j \Delta l_k \rangle$. For the exact $\Delta l_i = l_i (u_0 + \Delta u ) - l_i (u_0 )$, such a correlator in the coordinate representation would be a power series in the propagator, which, when the propagator is small, that is, at large distances, reduces to the main term, the propagator itself.

\section*{Acknowledgments}

The present work was supported in part by the Ministry of Education and Science of the Russian Federation.

\end{document}